\documentclass[letterpaper]{article} 
\usepackage{aaai2027}  
\usepackage[hyphens]{url}  
\usepackage{graphicx} 
\usepackage{natbib}  
\usepackage{caption} 
\usepackage{algorithm}

\usepackage{newfloat}
\usepackage{listings}
\DeclareCaptionStyle{ruled}{labelfont=normalfont,labelsep=colon,strut=off} 
\floatstyle{ruled}
\newfloat{listing}{tb}{lst}{}
\floatname{listing}{Listing}

\usepackage{booktabs}

\usepackage{microtype}
\usepackage{complexity}
\usepackage{siunitx}
\usepackage{algpseudocode}
\usepackage{algpseudocodex}

\algrenewcommand\algorithmicindent{1em}
\usepackage{amsthm}
\usepackage{amssymb}
\usepackage{mathtools}
\usepackage{cleveref}
\crefname{table}{Table}{Tables}
\crefname{prop}{Proposition}{Proposition}
\usepackage{makecell}
\usepackage{appendix}
\usepackage{tikz}
\usepackage{thm-restate}
\usepackage{enumitem}
\usepackage{subcaption}
\usetikzlibrary{decorations.pathreplacing}
\usetikzlibrary{positioning,shapes.geometric,fit}
\usetikzlibrary{arrows.meta,shapes.misc}
\usetikzlibrary{intersections}
\theoremstyle{plain}
\newtheorem{theorem}{Theorem}

\newtheorem{proposition}[theorem]{Proposition}
\newtheorem{corollary}[theorem]{Corollary}
\newtheorem{claim}{Claim}
\crefname{proposition}{Proposition}{Propositions}
\crefname{claim}{Claim}{Claims}

\RequirePackage{xspace}
\newcommand{\n}[1]{{#1}\xspace}  
\DeclareMathOperator*{\dist}{\textsf{dist}}

\newcommand{\prbname}{\n{\textsc{$r$CUMAPF}}}
\newcommand{\oneprbname}{\n{\textsc{1CUMAPF}}}
\newcommand{\mapf}{\n{\textsc{MAPF}}}

\renewcommand{\SAT}{\n{\textsc{SAT}}}
\newcommand{\TRUE}{\texttt{TRUE}}
\newcommand{\FALSE}{\texttt{FALSE}}

\newcommand{\indG}[2]{#1[#2]}
\newcommand{\nei}[2]{N_{#1}(#2)}
\newcommand{\neicl}[2]{N_{#1}[#2]}

\newcommand{\conf}[1]{\mathcal{Q}_{#1}}

\usepackage{threeparttable}
\usepackage{adjustbox}

\makeatletter
\def\input@path{{../}}
\makeatother

\title{Even Loose Proximity Constraints Make \\ Makespan-optimal Unlabeled Multi-agent Pathfinding Hard}

\author{
    Takahiro Suzuki\textsuperscript{\rm 1,2},
    Keisuke Okumura\textsuperscript{\rm 2}
}
\affiliations{
    \textsuperscript{\rm 1}Graduate School of Information Sciences, Tohoku University, Japan \\
    \textsuperscript{\rm 2}National Institute of Advanced Industrial Science and Technology (AIST), Japan\\
    takahiro.suzuki.q4@dc.tohoku.ac.jp, okumura.k@aist.go.jp
}

\nocopyright
\begin{document}

\maketitle

\begin{abstract}
We study \textsc{$r$-Connected Unlabeled Multi-agent Pathfinding} (\prbname), a variant of \mapf where interchangeable agents have a proximity range $r\in \mathbb{N}$, and agents must form a single group in a given graph $G$ through limited-range communication at all times.
The problem arises in fundamental applications such as self-reconfiguration and platooning, while standard \mapf does not capture proximity constraints.
For the problem of finding a makespan-optimal solution, prior work shows \NP-hardness for $r=1$ and polynomial-time solvability when $r$ is sufficiently large; the complexity for the intermediate case $2\le r$ is largely unknown.
To narrow this gap, we show that finding a solution to \prbname with makespan at most $2$ is \NP-hard for $2\le r= O(|V|^{1-\varepsilon})$, where $0<\varepsilon\le 1$ is a fixed constant and $|V|$ is the number of vertices.
As a consequence, we show that the labeled analogue of \prbname is also \NP-hard under the same condition.
\end{abstract}

\section{Introduction}
Research on \emph{Multi-Agent Pathfinding (\mapf)} has attracted considerable attention in the last decade, driven by the growing demand for modern cooperative multi-robot systems.
\mapf aims to find a set of collision-free paths on a given graph $G=(V,E)$, or a \emph{plan}, from initial positions to target positions for multiple agents.
Standard \mapf assumes discrete time, allows each agent to remain at its current vertex or move to an adjacent vertex at each time step, and requires plans to avoid the following two types of \emph{collisions}: \emph{(i)} vertex collisions, where more than one agent occupies the same vertex $v\in V$ at the same time, and \emph{(ii)} swap collisions, where two agents traverse an edge $e\in E$ at the same time.

In response to diverse requirements, recent work proceeds to study variants of \mapf (see, e.g.,~\cite{Stern19,Zahradka26}), since a prescribed standard formulation may not always capture constraints required in practical applications, such as safety~\cite{mapf:Atzmon20,drone:honig17}, uncertainty~\cite{uncertainMAPF/0001KK17}, and limited inter-agent communication range (i.e., integrity)~\cite{CMAPF/MaoudjC24}.
This paper also studies one of the variants, namely the proximity-constrained variant, since such constraints arise in collaborative work in space~\cite{Aerospace/Liu2026}, autonomous exploration at disaster sites~\cite{Disaster/TranzattoDBCKMPWZKRCHPO24}, and scenarios involving agents with a finite sensing range~\cite{sensing/CarliSG24}.
For $r\in\mathbb{N}$, the \emph{proximity constraint} requires the occupied vertices to induce a connected subgraph of $G^r$ at every time step, where $G^r$ joins vertices at distance at most $r$ in $G$.
We say $r$ is the \emph{proximity range}.
One can see that this proximity constraint naturally captures the requirement that any two agents can exchange information through multi-hop communication. 

Given that finding optimal solutions for labeled MAPF is \NP-hard regarding various metrics~\cite{Yu13hard}, our complextiy study focuses on the \emph{unlabeled} variant, which admits a polynomial-time makespan-optimal algorithm~\cite{Yu13unlabeled}, where \emph{makespan} denotes the length of the plan. 
In the unlabeled setting, every pair of agents is interchangeable, and the positions of agents are given by a set of vertices.
Together with the proximity constraint, we study \textsc{$r$-Connected Unlabeled Multi-agent Pathfinding} (\prbname), which seeks a plan between given initial and target vertex sets with minimum makespan.
See \Cref{sec:prelim} for a formal definition of our problem.

Although the importance of \prbname is underscored by a wide range of applications---such as autonomous platooning~\cite{Diaz21} and programmable matter~\cite{Hinnenthal24}---only limited complexity results are known for this problem.
For the extremal cases of makespan optimization, \citet{Fekete21} show that deciding whether there is a plan for \oneprbname with makespan at most $2$ is \NP-complete,
and the case $r=D(G)$, where $D(G)$ is the diameter of $G$, can be reduced to unlabeled \mapf~\cite{Yu13unlabeled}, which is known to be solvable in polynomial time.
Moreover, existing suboptimal algorithms~\cite{Fekete21,Suzuki26ICAPS} only consider \oneprbname, and they cannot be applied for $r\ge 2$, which evidently reveals the challenge of extending the results to the general case.
Reflecting these results, a natural question arises from the viewpoint of computational complexity: \emph{does \prbname for the case $2\le r<D(G)$ admit a polynomial-time algorithm, or is it \NP-hard?}

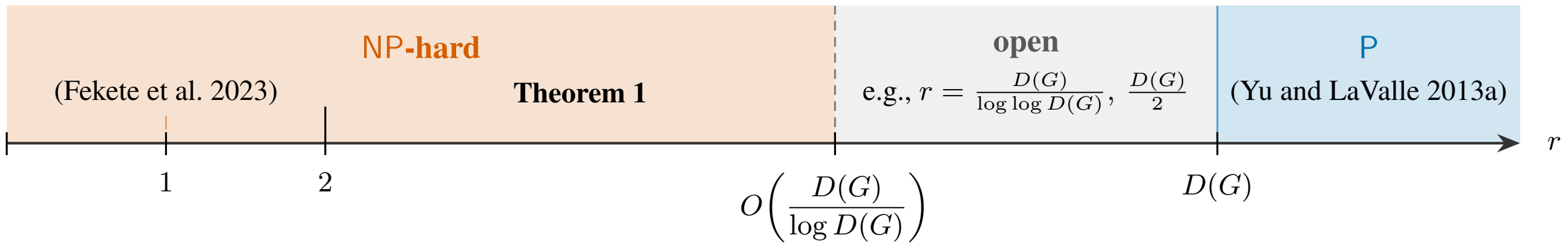
\begin{figure*}[t]
    \centering
    \begin{tikzpicture}[x=0.0098\textwidth,y=1cm,font=\small]
        \definecolor{rangehard}{HTML}{D55E00}
        \definecolor{rangepoly}{HTML}{0072B2}

        \fill[rangehard!18] (0,0) rectangle (52,1.5);
        \fill[black!6] (52,0) rectangle (76,1.5);
        \fill[rangepoly!18] (76,0) rectangle (95,1.5);
        \draw[gray,densely dashed,line width=0.6pt] (52,0) -- (52,1.5);
        \draw[rangepoly!65,line width=0.6pt] (76,0) -- (76,1.5);
        \draw[black!80,line width=0.8pt,-{Stealth[length=2.5mm,width=2mm]}]
            (0,0) -- (95,0);

        \draw[black,line width=0.6pt] (0,0.12) -- (0,-0.13);
        \foreach \x/\ticklabel in {10/1,20/2,52/{O\!\left(\dfrac{D(G)}{\log D(G)}\right)},76/D(G)} {
            \draw[black,line width=0.6pt] (\x,0.12) -- (\x,-0.13);
            \node[anchor=north] at (\x,-0.2) {$\ticklabel$};
        }
        \draw[black,line width=0.6pt] (20,0.12) -- (20,0.4);
        \node[anchor=west] at (96,0) {$r$};

        \node[text=rangehard,font=\bfseries] at (26,1.05) {\NP-hard};
        \node at (10,0.55) {\cite{Fekete21}};
        \draw[rangehard!65,line width=0.5pt] (10,0.15) -- (10,0.3);
        \node at (36,0.55) {\textbf{\Cref{thm:CUMAPFhard}}};
        \node[font=\bfseries,text=black!65] at (64,1.05) {open};
        \node at (64,0.55) {e.g., $r=\frac{D(G)}{\log\log D(G)},\,\frac{D(G)}{2}$};
        \node[text=rangepoly,font=\bfseries] at (85.5,1.05) {\P};
        \node at (85.5,0.55) {\cite{Yu13unlabeled}};
    \end{tikzpicture}
    \caption{Complexity of makespan-optimal \prbname by proximity range relative to $D(G)$.}
    \label{fig:complexity-range}
\end{figure*}

\subsection*{Our Contribution}
We partially answer this question by proving the following theorem:

\begin{theorem}\label{thm:CUMAPFhard}
    For every fixed constant $0<\varepsilon\le 1$ and every integer $r\ge 2$ satisfying $r=O(|V|^{1-\varepsilon})$, it is \NP-hard to find a plan for \prbname with a makespan of at most $2$, even if $G$ has maximum degree $3$.
\end{theorem}

Note that an $O(|V|^\varepsilon)$ gap remains between the tractable and intractable cases.
In terms of the diameter $D(G)$, however, this gap is only $O(\log D(G))=O(\log |V|)$, since the reduction establishes \NP-hardness for $r=O(D(G)/\log D(G))$ and the problem is polynomial-time solvable for $r\ge D(G)$.
Moreover, the reduction yields a running-time lower bound depending on $\varepsilon$ under a standard complexity assumption, as well as \NP-hardness for the labeled variant of \prbname, indicating that imposing proximity constraints makes the problem challenging significantly.
\Cref{fig:complexity-range} summarizes the known complexity regimes for \prbname.

\paragraph{Open problems and future work.}
Although our theorem establishes a large part of the complexity classification, there is still a gap between tractable and intractable cases.
In particular, the complexity for the case $r=\omega (D(G)/\log D(G))$ is still unknown.
Moreover, our reduction does not rule out the existence of a subexponential-time algorithm depending on $\varepsilon$; developing such a complementary algorithm could be a direction for future work.

\paragraph{Paper organization.}
In the following, we define notations and the problem statement (\Cref{sec:prelim}), introduce related work (\Cref{sec:CUMAPFrelated}), and prove \Cref{thm:CUMAPFhard}  (\Cref{sec:hard}) together with algorithmic consequences (\Cref{sec:hardnessconseq}).

\section{Preliminaries}\label{sec:prelim}
\subsection{Notation}
Let $G$ be a simple, undirected, unweighted, and finite connected graph. 
We write the vertex set and the edge set of $G$ as $V(G)$ and $E(G)$, respectively.
For a vertex $v \in V(G)$, the \emph{open neighborhood} is $\nei{G}{v} = \{w \in V(G) \mid vw \in E(G)\}$.
The maximum degree of $G$ is $\Delta(G) = \max_{v \in V(G)} |\nei{G}{v}|$.
For a vertex subset $S \subseteq V(G)$, we denote the subgraph induced by $S$ as $\indG{G}{S}$. 
We use $[i,j]$ to denote the set $\{i,i+1,\dots, j-1,j\}$ of consecutive integers for $i\le j$.
The distance $\dist(u,v)$ is the number of edges in a shortest $u$--$v$ path.
A path $P=(p_i)_{i=0}^k$ has length $k$ and vertex set $V(P)=\{p_i\mid i\in [0,k]\}$.

\subsection{Problem Definition}\label{sec:prbdef}
Let $G$ be a graph and $A = [1,n]$ be a set of $n$ agents. 
A \emph{configuration} $\mathcal{Q} = \{q_1, \dots, q_n\} \subseteq V(G)$ is a set of vertices.
\paragraph{Proximity constraint.}
To formally define our proximity constraint (also appearing in~\cite{CUMAPF/FioravantesKKMO26}) on $G$, we introduce an auxiliary graph $G^r$, defined by adding an edge $uv$ for every pair of vertices $u,v$ such that $\dist(u,v)\le r$.
A configuration $\mathcal{Q}$ is \emph{$r$-connected} if $\indG{G^r}{\mathcal{Q}}$ is connected. 
Moreover, for a vertex $u$ of $G$, we say that a vertex $v$ is an \emph{$r$-neighbor} of $u$ when $v\in\neicl{G^r}{u}$, that is, $\dist(u,v)\le r$ in $G$.

\paragraph{Problem formulation.}
Given two distinct sets of vertices $S, T \subseteq V(G)$ with $|S|=|T|=n$, the goal of \prbname is to find a finite sequence of configurations (a \emph{plan}) $\Pi = [S=\conf{0}, \conf{1}, \ldots, \conf{t^{\ast}}=T]$.
Throughout the plan, agents can only move to adjacent (or the same) vertices and must avoid vertex and swap conflicts while maintaining $r$-connectivity.
Formally, for a fixed integer $1\le r\le |V(G)|$, a \emph{valid} plan $\Pi$ for \prbname satisfies the following conditions:
\begin{enumerate}
   \item $\conf{0} = S\land \conf{t^{\ast}} = T$.
   \item $\forall i \in A$, $\forall k \in [1, t^{\ast}]: \mathcal{Q}_{k}[i] \in \nei{G}{\mathcal{Q}_{k-1}[i]}\cup \{\mathcal{Q}_{k-1}[i]\}$. (\emph{reachability})
   \item $\forall k \in [0,t^{\ast}]$, $ \forall i, j \in A$, $i \neq j: \mathcal{Q}_k[i] \neq \mathcal{Q}_k[j]$. (avoidance of \emph{vertex collision})
   \item $\forall k \in [1,t^{\ast}]$, $\forall i, j \in A$, $i \neq j: (\conf{k}[i],\conf{k}[j])\ne (\conf{k-1}[j],\conf{k-1}[i])$.  (avoidance of \emph{swap collision})
   \item $\forall k \in [0,t^{\ast}]: \conf{k}$ is $r$-connected.
\end{enumerate}
The makespan of a plan $\Pi$ is defined by $t^{\ast}$.
A valid plan for the labeled analogue of \prbname, denoted by \textsc{$r$CMAPF}, is given by the same condition, except that the initial and target configurations are input as lists of vertices.

\section{Related Work}\label{sec:CUMAPFrelated}
Beyond classical \mapf, practical settings motivate additional requirements, including alternative collision definitions~\cite{Li19large,Surynek19capacity,Suzuki26distance} and constraints on configurations~\cite{Li20,Tateo2018,drone:honig17}.
Among these, several studies consider constraints motivated by limited communication between agents and integration:
\citet{Tateo2018} study communication constraints in a model where physical movement and communication are represented separately; 
\citet{Li20} introduce a soft constraint for maintaining a specific formation; 
and \citet{Otaki20} address a platooning scenario with an objective that considers air resistance.
Recent work also considers planning with graph-based proximity constraints, as follows.
For \prbname, \citet{Fekete21} show that deciding whether a plan with makespan $2$ exists for \oneprbname is \NP-complete, even on empty 2D grids, and provide polynomial-time constant-factor approximation algorithms for such grids.
For \textsc{$r$CMAPF}, \citet{CUMAPF/FioravantesKKMO26} provide parameterized complexity results, and \citet{CMAPF/FeketeKRSS24} establish makespan lower bounds and provide constant-factor approximation algorithms for restricted cases.
In contrast, our work considers the unlabeled setting with an arbitrary proximity range $r$.

\section{Proof of \Cref{thm:CUMAPFhard}}\label{sec:hard}

We prove the hardness by reducing the Boolean satisfiability problem (\SAT) to \prbname.
Let $\phi$ be a CNF formula with clause set $C=\{c_1,c_2,\dots,c_m\}$ and variable set $X=\{x_1,x_2,\dots,x_n\}$.
\SAT is the problem of deciding whether $\phi$ is satisfiable, given a CNF formula $\phi$.
It is known that \SAT is \NP-complete, even when each clause contains
exactly three literals and each variable occurs exactly twice positively
and twice negatively~\cite{SAT/Darmann21}.
We assume that such a formula is given.

\subsection{Construction}
\begin{figure*}
	\centering
    \begin{minipage}{0.4\textwidth}
    \centering
    \begin{subfigure}[c]{\textwidth}
    \centering
    \resizebox{\linewidth}{!}{\definecolor{markedblue}{HTML}{2368A2}
\definecolor{agentvariable}{HTML}{D73027}

\tikzset{
  figone edge/.style={draw=black,line width=2.25pt,line cap=round},
  figone rminusone/.style={figone edge,dash pattern=on 3pt off 4pt},
  figone vertex/.style={circle,draw=black,line width=2.25pt,fill=white,
                        minimum size=0.757cm,inner sep=0pt},
  figone occupied/.style={figone vertex,fill=agentvariable},
  figone ring/.style={draw=markedblue,line width=2.25pt},
  figone label/.style={font=\fontsize{24pt}{26pt}\selectfont,inner sep=0pt}
}

\begin{tikzpicture}[x=0.02cm,y=-0.02cm]
  \coordinate (fig1-aL) at (128.76,414.67);
  \coordinate (fig1-aR) at (243.42,414.67);
  \coordinate (fig1-aT) at (185.42,480.67);
  \coordinate (fig1-aM) at (185.42,554.04);
  \coordinate (fig1-aB) at (185.42,627.40);

  \coordinate (fig1-bL) at (349.98,414.67);
  \coordinate (fig1-bR) at (464.64,414.67);
  \coordinate (fig1-bT) at (406.64,480.67);
  \coordinate (fig1-bM) at (406.64,554.04);
  \coordinate (fig1-bB) at (406.64,627.40);

  \coordinate (fig1-cL) at (571.09,414.67);
  \coordinate (fig1-cR) at (685.75,414.67);
  \coordinate (fig1-cT) at (627.75,480.67);
  \coordinate (fig1-cM) at (627.75,554.04);
  \coordinate (fig1-cB) at (627.75,627.40);

  \coordinate (fig1-dL) at (792.17,413.82);
  \coordinate (fig1-dR) at (906.83,413.82);
  \coordinate (fig1-dT) at (848.83,479.82);
  \coordinate (fig1-dM) at (848.83,553.18);
  \coordinate (fig1-dB) at (848.83,626.55);

  \foreach \l/\t/\r/\m/\b in {
    fig1-aL/fig1-aT/fig1-aR/fig1-aM/fig1-aB,
    fig1-bL/fig1-bT/fig1-bR/fig1-bM/fig1-bB,
    fig1-cL/fig1-cT/fig1-cR/fig1-cM/fig1-cB,
    fig1-dL/fig1-dT/fig1-dR/fig1-dM/fig1-dB}
    \draw[figone edge] (\l)--(\t)--(\r) (\t)--(\m)--(\b);

  \draw[figone rminusone] (fig1-aB)--(fig1-bM);
  \draw[figone rminusone] (fig1-bB)--(fig1-cM);
  \draw[figone rminusone] (fig1-cB)--(fig1-dM);

  \foreach \v in {fig1-aT,fig1-aB,fig1-bT,fig1-bB,
                    fig1-cT,fig1-cB,fig1-dT,fig1-dB}
    \draw[figone ring] (\v) circle[radius=0.5905cm];

  \foreach \v in {fig1-aL,fig1-aR,fig1-aM,
                    fig1-bL,fig1-bR,fig1-bM,
                    fig1-cL,fig1-cR,fig1-cM,
                    fig1-dL,fig1-dR,fig1-dM}
    \node[figone vertex] at (\v) {};

  \foreach \v in {fig1-aT,fig1-aB,fig1-bT,fig1-bB,
                    fig1-cT,fig1-cB,fig1-dT,fig1-dB}
    \node[figone occupied] at (\v) {};

  \node[figone label,above=13pt] at (fig1-aL) {$u_{1,T}$};
  \node[figone label,above=13pt] at (fig1-aR) {$u_{1,F}$};
  \node[figone label,left=24pt] at (fig1-aT) {$u_{1,2}$};
  \node[figone label,left=18pt] at (fig1-aM) {$u_{1,1}$};
  \node[figone label,left=24pt] at (fig1-aB) {$u_{1,0}$};
  \foreach \prefix in {b,c,d}{
    \node[figone label,above=13pt] at (fig1-\prefix L) {$T$};
    \node[figone label,above=13pt] at (fig1-\prefix R) {$F$};
  }
\end{tikzpicture}}
    \caption{}
    \label{fig:var}
  \end{subfigure}
  \smallskip
      \begin{subfigure}[c]{0.75\textwidth}
    \centering
    \resizebox{\linewidth}{!}{\definecolor{markedblue}{HTML}{2368A2}
\definecolor{agentclause}{HTML}{F28E2B}
\tikzset{
  edge/.style={draw=black,line width=2.25pt,line cap=round},
  rminusone edge/.style={edge,dash pattern=on 3pt off 4pt},
  vertex/.style={circle,draw=black,line width=2.25pt,fill=white,
                 minimum size=0.757cm,inner sep=0pt},
  occupied vertex/.style={vertex,fill=agentclause},
  marked ring/.style={draw=markedblue,line width=2.25pt},
  clause label/.style={font=\fontsize{24pt}{26pt}\selectfont,inner sep=0pt}
}

\begin{tikzpicture}[x=0.02cm,y=-0.02cm]
  \coordinate (aT) at (185.42,471.78);
  \coordinate (aM) at (185.42,551.00);
  \coordinate (aB) at (185.42,627.40);
  \coordinate (bT) at (406.64,471.78);
  \coordinate (bM) at (406.64,551.00);
  \coordinate (bB) at (406.64,627.40);
  \coordinate (cT) at (627.75,471.78);
  \coordinate (cM) at (626.88,551.57);
  \coordinate (cB) at (626.72,627.40);

  \coordinate (tAB) at (295.08,471.78);
  \coordinate (tBC) at (515.31,471.78);
  \coordinate (bAB) at (295.08,627.40);
  \coordinate (bBC) at (515.31,627.40);

  \foreach \t/\m/\b in {aT/aM/aB,bT/bM/bB,cT/cM/cB}
    \draw[edge] (\t)--(\m)--(\b);
  \draw[edge] (aT)--(tAB) (bT)--(tBC)
              (aB)--(bAB) (bB)--(bBC);
  \draw[rminusone edge] (tAB)--(bT) (tBC)--(cT)
                       (bAB)--(bB) (bBC)--(cB);

  \foreach \v in {aB,bB,cB}
    \draw[marked ring] (\v) circle[radius=0.5698cm];

  \foreach \v in {aM,bM,cM,aB,bB,cB,tAB,tBC,bAB,bBC}
    \node[vertex] at (\v) {};
  \foreach \v in {aT,bT,cT}
    \node[occupied vertex] at (\v) {};

  \node[clause label,left=18pt] at (aT) {$p_{i,0}$};
  \node[clause label,left=18pt] at (aM) {$p_{i,1}$};
  \node[clause label,left=24pt] at (aB) {$p_{i,2}$};
  \node[clause label,above=13pt] at (aT) {$1$};
  \node[clause label,above=13pt] at (bT) {$2$};
  \node[clause label,above=13pt] at (cT) {$3$};
\end{tikzpicture}}
    \caption{}
    \label{fig:clause}
  \end{subfigure}
  \smallskip
      \begin{subfigure}[c]{0.3\textwidth}
    \centering
    \resizebox{\linewidth}{!}{\definecolor{markedblue}{HTML}{2368A2}
\definecolor{agentconnection}{HTML}{7A0177}
\tikzset{
  edge/.style={draw=black,line width=2.25pt,line cap=round},
  rminusone edge/.style={edge,dash pattern=on 3pt off 4pt},
  vertex/.style={circle,draw=black,line width=2.25pt,fill=white,
                 minimum size=0.757cm,inner sep=0pt},
  occupied vertex/.style={vertex,fill=agentconnection},
  marked ring/.style={draw=markedblue,line width=2.25pt},
  connection label/.style={font=\fontsize{30pt}{32pt}\selectfont,inner sep=0pt}
}

\begin{tikzpicture}[x=0.02cm,y=-0.02cm]
  \coordinate (top) at (500,180);
  \coordinate (upper) at (500,280);
  \coordinate (center) at (420,360);
  \coordinate (right) at (580,360);
  \coordinate (lower) at (500,440);
  \coordinate (bottom) at (500,540);

  \draw[edge] (top)--(upper)--(center)--(right);
  \draw[rminusone edge] (center)--(lower);
  \draw[edge] (lower)--(bottom);

  \foreach \v in {right,lower}
    \draw[marked ring] (\v) circle[radius=0.5698cm];

  \foreach \v in {top,center,right,bottom}
    \node[vertex] at (\v) {};
  \foreach \v in {upper,lower}
    \node[occupied vertex] at (\v) {};

  \node[connection label,above=13pt] at (top)    {$w_1$};
  \node[connection label,left=18pt]  at (upper)  {$w_2$};
  \node[connection label,left=18pt]  at (center) {$w_3$};
  \node[connection label,right=24pt] at (right)  {$w_4$};
  \node[connection label,right=24pt] at (lower)  {$w_5$};
  \node[connection label,below=13pt] at (bottom) {$w_6$};
\end{tikzpicture}}
    \caption{}
    \label{fig:conn}
  \end{subfigure}
  \hfill
      \begin{subfigure}[c]{0.35\textwidth}
    \centering
    \resizebox{\linewidth}{!}{\definecolor{markedblue}{HTML}{2368A2}
\definecolor{agentoccurrence}{HTML}{FF4F9A}
\tikzset{
  edge/.style={draw=black,line width=2.25pt,line cap=round},
  rminusone edge/.style={edge,dash pattern=on 3pt off 4pt},
  vertex/.style={circle,draw=black,line width=2.25pt,fill=white,
                 minimum size=0.757cm,inner sep=0pt},
  occupied vertex/.style={vertex,fill=agentoccurrence},
  marked ring/.style={draw=markedblue,line width=2.25pt},
  occurrence label/.style={font=\fontsize{32pt}{34pt}\selectfont,inner sep=0pt}
}

\begin{tikzpicture}[x=0.02cm,y=-0.02cm]
  \coordinate (top2) at (500,100);
  \coordinate (top3) at (500,180);
  \coordinate (upperRed) at (500,260);

  \coordinate (upperLeft) at (400,340);
  \coordinate (upperRight) at (600,340);
  \coordinate (lowerRed) at (700,420);
  \coordinate (lowerBranch) at (600,420);
  \coordinate (lowerLeft) at (520,500);
  \coordinate (lowerRight) at (680,500);

  \draw[rminusone edge] (top2)--(top3);
  \draw[edge] (top3)--(upperRed);
  \draw[rminusone edge] (upperLeft)--(upperRed);
  \draw[edge] (upperRed)--(upperRight)--(lowerRed);
  \draw[edge] (upperRight)--(lowerBranch);
  \draw[rminusone edge] (lowerLeft)--(lowerBranch)--(lowerRight);

  \foreach \v in {upperRed,lowerRed}
    \draw[marked ring] (\v) circle[radius=0.5698cm];

  \foreach \v in {top2,top3,upperLeft,upperRight,lowerBranch,lowerLeft,lowerRight}
    \node[vertex] at (\v) {};
  \foreach \v in {upperRed,lowerRed}
    \node[occupied vertex] at (\v) {};

  \node[occurrence label,left=18pt]  at (upperRed) {$a_j$};
  \node[occurrence label,right=18pt] at (lowerRed) {$b_j$};
  \node[occurrence label,left=18pt]  at (top3) {$a_{p,j}$};
  \node[occurrence label,right=18pt] at (upperRight) {$b_{a,j}$};
\end{tikzpicture}}
    \caption{}
    \label{fig:occ}
  \end{subfigure}
  \end{minipage}
  \begin{minipage}{0.55\textwidth}
  \begin{subfigure}[c]{\textwidth}
  \resizebox{\linewidth}{!}{\input{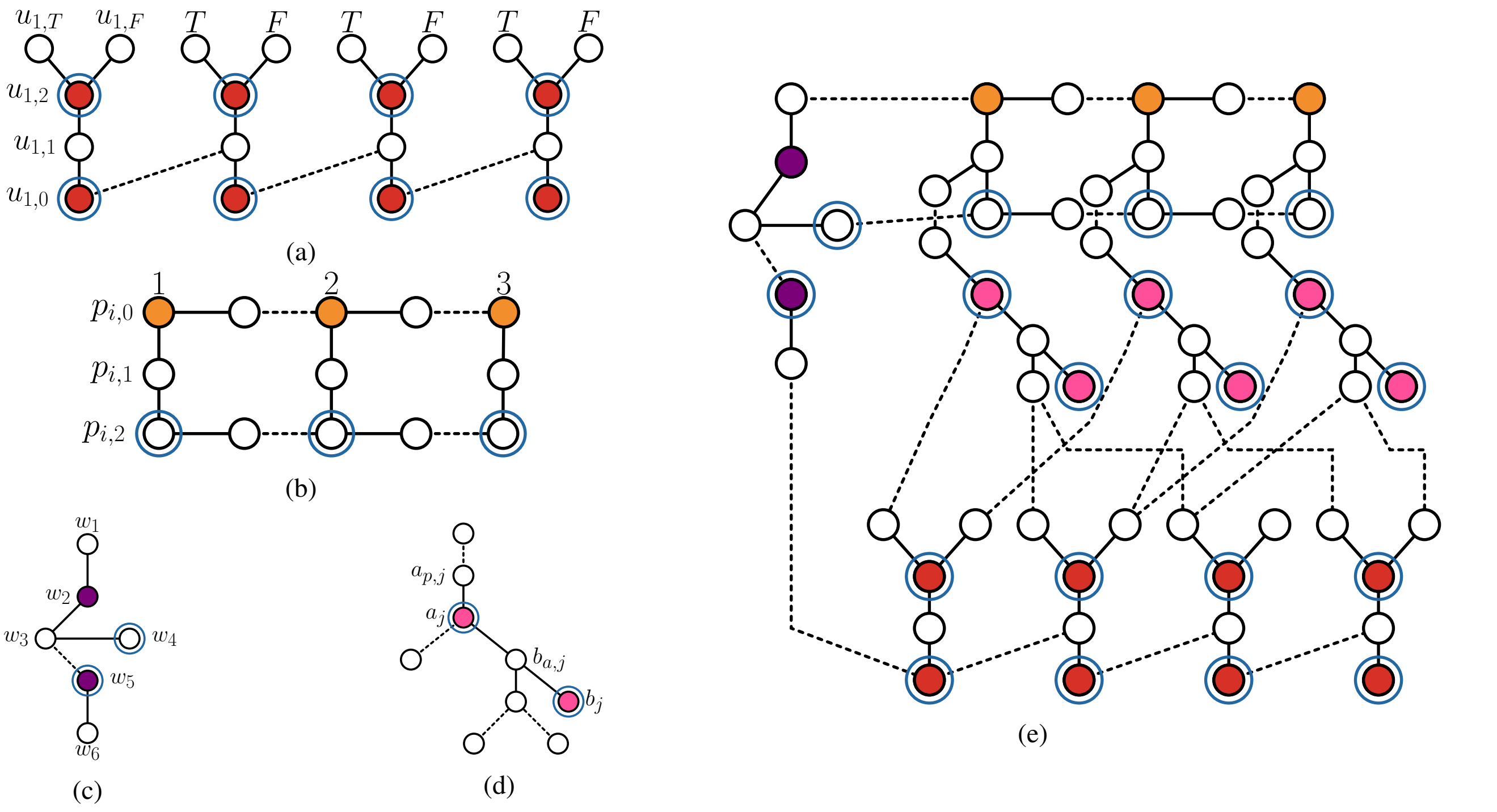}}
  \caption{}
  \end{subfigure}
  \end{minipage}
  \begin{minipage}{0.45\textwidth}
      
  \end{minipage}
	\caption{Illustrations of the construction when $r=2$ and $\phi = (x_1\lor x_2\lor x_3)\land (\neg x_1 \lor \neg x_2\lor  x_4)\land (\neg x_2 \lor x_3\lor \neg x_4)$. \emph{(a):} variable gadgets, \emph{(b):} clause gadgets, \emph{(c):} the connection gadget, and \emph{(d):} occurrence gadgets. \emph{(e):} The entire graph $G$ and two configurations $S$, $T$. Filled vertices indicate $S$, with colors identifying their gadgets, and blue circles indicate $T$. Solid lines represent single edges, whereas dotted lines represent paths of length $r-1$.}
	\label{fig:gadgets}
\end{figure*}
Now we construct an instance $(G=(V,E), S,T)$ for a fixed integer $r$ from a CNF formula $\phi$.
The graph $G$ consists of the following components: \emph{variable gadgets}, \emph{clause gadgets}, a \emph{connection gadget}, and \emph{occurrence gadgets}.
We now introduce these components one by one.
As we construct the gadgets, we also specify the vertex subsets $S$ and $T$.
Illustrations of our gadgets and resulting instances can be seen in \Cref{fig:gadgets}.

\paragraph{Variable gadget (\Cref{fig:var}).} 
We first introduce $n$ paths $U_1,U_2,\ldots,U_n$, where $U_i = (u_{i,j})_{j=0}^r$ for each $i\in[1,n]$.
Then, we add the set of vertices $\{u_{i,T}, u_{i,F}\mid i\in [1,n]\}$ and edges $\{u_{i,T}u_{i,r}, u_{i,F}u_{i,r}\mid i\in [1,n]\}$.
Last, we connect these gadgets by paths: for every $i\in [1,n-1]$, we connect $u_{i,0}$ and $u_{i+1,1}$ by a single path, so that $\dist(u_{i,0},u_{i+1,1})=r-1$.
For the variable gadgets, we include $\{u_{i,0}, u_{i,r}\mid i\in [1,n]\}$
in $S$ and $T$.
\paragraph{Clause gadget (\Cref{fig:clause}).} 
We introduce $m$ paths $P_{1},P_{2},\ldots,P_{m}$ of length $2$, where $V(P_{i})=\{p_{i,0},p_{i,1},p_{i,2}\}$.
For every $i\in[1,m-1]$, we connect $p_{i,0}$ to $p_{i+1,0}$ and $p_{i,2}$ to $p_{i+1,2}$ by paths so that
$
\dist(p_{i,0},p_{i+1,0})
=
\dist(p_{i,2},p_{i+1,2})
=
r$.
We include $\{p_{i,0}\mid i\in [1,m]\}$ in $S$ and $\{p_{i,2}\mid i\in [1,m]\}$ in $T$.
\paragraph{Connection gadget (\Cref{fig:conn}).}
We introduce six vertices $w_1,w_2,\ldots,w_6$ and add the edges $w_1w_2$, $w_2w_3$, $w_3w_4$, and $w_5w_6$.
We introduce paths of length $r-1$ between $w_3$ and $w_5$, between $p_{1,0}$ and $w_1$, between $w_4$ and $p_{1,2}$, and between $w_6$ and $u_{1,0}$.
We include $w_2$ and $w_5$ in $S$, and $w_4$ and $w_5$ in $T$.
\paragraph{Occurrence gadget (\Cref{fig:occ}).}
Assume that a clause $c_j$ consists of $3$ literals $l_{1,j}, l_{2,j},l_{3,j}$.
For each $i\in [1,3]$, let $v_{i,j}^\ast=u_{k,T}$ (resp. $v_{i,j}^\ast=u_{k,F}$) if $l_{i,j}$ is a positive literal (resp. negative literal) of variable $x_k$.
We introduce two vertices $a_j$ and $b_j$, a path of length $r+1$ between $p_{j,1}$ and $a_j$, a path of length $r-1$ between $a_j$ and $v_{1,j}^\ast$, and a path of length $r$ between $a_j$ and $b_j$.
Let $a_{p,j}$ be the neighbor of $a_j$ on the introduced $p_{j,1}$--$a_j$ path, and let $b_{a,j}$ be the neighbor of $b_j$ on the introduced $a_j$--$b_j$ path.
Finally, we add a new vertex adjacent to $b_{a,j}$ and connect this new vertex to each of $v_{2,j}^\ast$ and $v_{3,j}^\ast$ by a path of length $r-1$.
We include $\{a_{j}, b_{j}\}$ in $S$ and $T$.

We can easily see that the maximum degree of $G$ is 3, and both $S$ and $T$ are $r$-connected (see \Cref{fig:gadgets,fig:conf1}). The following claim completes the construction.
\begin{claim}
	The resulting graph $G$ has a size polynomial in $n$.
\end{claim}
\begin{proof}
	We first count the number of vertices of each gadget, using $r$ and $n$.
	\begin{enumerate}
		\item \textbf{variable gadget:} there are $\Theta(n)$ paths of length $\Theta(r)$, and thus there are $O(rn)$ vertices.
		\item \textbf{clause gadget:} there are $m$ paths of constant length
        and $2(m-1)$ connecting paths of length $r$, and hence
        $\Theta(rm)$ vertices.
		\item \textbf{connection gadget:} the gadget consists of $\Theta(r)$ vertices.
		\item \textbf{occurrence gadget:} each gadget consists of $5$ paths of length $\Theta(r)$, so there are $\Theta(r)$ vertices per gadget.
	\end{enumerate}
	Hence, the number of vertices is $\Theta(rn)+\Theta(rm)+\Theta(r) + m\cdot \Theta(r) = \Theta(r(n+m))$.
	Since each variable occurs at most four times in $\phi$, the number of clauses $m$ is at most $4n=O(n)$.
    Thus, $|V(G)|=\Theta(rn)$.
    
Fix a constant $0<\epsilon\le 1$ and any integer $r\ge 2$ such that
$r=O\left(n^{(1-\epsilon)/\epsilon}\right)$.
Since $\epsilon$ is fixed, $r$ is polynomial in $n$, and hence the reduction runs in polynomial time.
Moreover, since $|V(G)|=\Theta(rn)$ and
$r^\epsilon=O\left(n^{1-\epsilon}\right)$, we have
$r=O\left(|V(G)|^{1-\epsilon}\right)$, as required.

\end{proof}

\subsection{Correctness}\label{sec:correctness}
\begin{figure*}[t]
	\centering
	\input{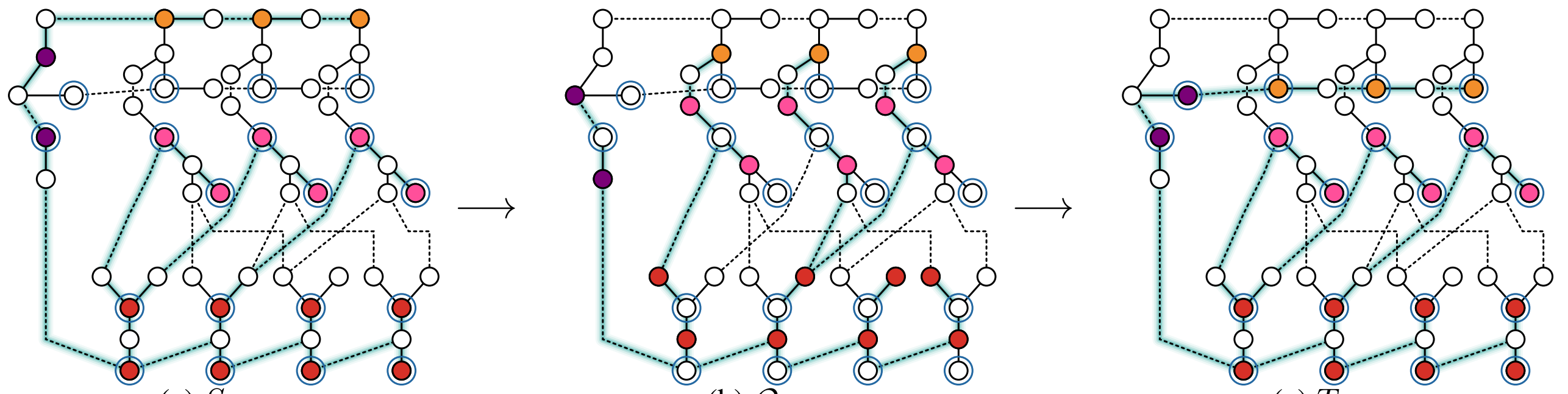}
	\caption{The plan $\langle S, \conf{1}, T\rangle$, where $\phi = (x_1\lor x_2\lor x_3)\land (\neg x_1 \lor \neg x_2\lor  x_4)\land (\neg x_2 \lor x_3\lor \neg x_4)$, and $\boldsymbol{x}=(\TRUE, \FALSE, \FALSE, \TRUE)$. Fill colors identify the agents' gadgets, and blue circles indicate $T$. Solid lines represent single edges, whereas dotted lines represent paths of length $r-1$. The green markings certify the $r$-connectivity of the three configurations.}
	\label{fig:conf1}
\end{figure*}

We show that there is a valid plan $(S=\conf{0}, \conf{1}, \conf{2}=T)$ of length $2$ from $S$ to $T$ for \prbname if and only if there is a satisfying assignment of $\phi$.
For the sake of simplicity, we use the bijections $f_1:S\to \conf{1}$ and $f_2:S\to T$ such that $f_1(v)$ (resp. $f_2(v)$) is the vertex $\conf{1}[i]$ (resp. $\conf{2}[i]$) for an agent $i$ such that $\conf{0}[i]=v$.
\paragraph{If direction.}
Assume there is a satisfying assignment $\boldsymbol{x}\in \{\TRUE, \FALSE\}^n$ for $\phi$, where the $i$-th component corresponds to the value of $x_i$.
We prove this direction by explicitly giving the functions $f_1,f_2$, thereby obtaining a valid plan $(S=\conf{0}, \conf{1}, \conf{2}=T)$.
First, we set $f_1$ (and hence $\conf{1}$) as follows: 
\begin{enumerate}
		\item \textbf{variable gadget:} for every index $i\in [1,n]$,
        \[
        \begin{aligned}
        f_1(u_{i,0})&=u_{i,1},\\
        f_1(u_{i,r})&= \begin{cases}
            u_{i,T}  \text{ if } \boldsymbol{x}_i=\TRUE,\\
            u_{i,F}  \text{ if } \boldsymbol{x}_i=\FALSE.
        \end{cases}
        \end{aligned}
        \]
		\item \textbf{clause gadget:} for every index $i\in [1,m]$, we set $f_1(p_{i,0})=p_{i,1}$.
		\item \textbf{connection gadget:} We set $f_1(w_2)=w_3$ and $f_1(w_5)=w_6$.
		\item \textbf{occurrence gadget:} Since $\boldsymbol{x}$ is a satisfying assignment of $\phi$, there is at least one literal among $l_{1,j}, l_{2,j},l_{3,j}$ whose corresponding value is $\TRUE$.
		We define $f_1(a_j)=a_{p,j}$ and $f_1(b_j)=b_{a,j}$.
	\end{enumerate}
	We can verify that $\conf{1}$ is a configuration that is conflict-free, $r$-connected, and reachable from $S$.
    An illustration of $\conf{1}$ is shown in \Cref{fig:conf1}.
	
Next, we set $f_2$.
\begin{enumerate}
	\item \textbf{variable gadget:} for every index $i\in [1,n]$, $f_2(u_{i,0})=u_{i,0}, \qquad
	f_2(u_{i,r})= u_{i,r}$.
	\item \textbf{clause gadget:} for every index $i\in [1,m]$, we set $f_2(p_{i,0})=p_{i,2}$.
	\item \textbf{connection gadget:} We set $f_2(w_2)=w_4$ and $f_2(w_5)=w_5$.
	\item \textbf{occurrence gadget:} For every $j\in[1,m]$, we set $f_2(a_j)=a_j$ and $f_2(b_j)=b_j$.
\end{enumerate}
We can check that $\conf{2}=T$ and $\conf{2}$ is reachable from $\conf{1}$ with no conflict.

\paragraph{Only if direction.}
Suppose that there is a valid plan $(S=\conf{0}, \conf{1}, \conf{2}=T)$.
We again use the notation $f_1$ and $f_2$.
We start by restricting the form of such bijections.
\begin{proposition}\label{prop:restrictp}
	For every index $j\in [1,m]$, $f_2(p_{j,0})=p_{j,2}$; hence, $f_1(p_{j,0})=p_{j,1}$.
\end{proposition}
\begin{proof}
	Since $(S=\conf{0}, \conf{1}, \conf{2}=T)$ is a valid plan, it is clear that $\dist(v, f_2(v))\le 2$ must hold for every vertex $v\in S$.
	For every index $j\in [2,m]$, $p_{j, 2}$ is the only vertex that is included in $T$ and has distance at most $2$ from $p_{j,0}$; hence, $f_2(p_{j,0})=p_{j,2}$.
	For $p_{1,0}$, consider the function $f_2^{-1}:T\to S$.
	Then, $p_{1,0}$ is the only vertex that is included in $S$ and has distance at most $2$ from $p_{1,2}$; we can conclude $f_2^{-1}(p_{1,2})=p_{1,0}$, and hence $f_2(p_{1,0})=p_{1,2}$.
	Thus, we have $f_2(p_{j,0})=p_{j,2}$ for every index $j\in [1,m]$.
	Moreover, since $(p_{j,0}, p_{j,1}, p_{j,2})$ is the only path of length $2$ for every index $j\in [1,m]$, it must hold that $f_1(p_{j,0})=p_{j,1}$.
\end{proof}

\begin{proposition}\label{prop:restrictw}
	For the connection gadget, $f_2(w_2)=w_4$.
\end{proposition}
\begin{proof}
	It is clear that $\{v\in T\mid \dist(v,w_2)\le 2\}$ is equal to $\{w_4\}$ when $r\ge 3$ and $\{w_4,w_5\}$ when $r=2$; we are done for $r\ge 3$ by a similar argument to that in \Cref{prop:restrictp}.
    Hence, we focus on the case $r=2$.
	In this case, $f_2(w_2)\in \{w_4,w_5\}$.
	However, if $f_2(w_2)=w_5$, we must have $f_2(w_5)=w_4$, since $\{v\in S\mid \dist(v,w_4)\le 2\}=\{w_2,w_5\}$.
	Moreover, since $(w_2,w_3,w_5)$ and $(w_5,w_3,w_4)$ are the only possible paths for the agents placed on $w_2$ and $w_5$, respectively, there is a vertex collision in $\conf{1}$.
	This contradicts the fact that $(S=\conf{0}, \conf{1}, \conf{2}=T)$ is a valid plan, and hence $f_2(w_2)=w_4$.
\end{proof}
By \Cref{prop:restrictp}, to ensure the $r$-connectivity of $\conf{1}$, we must have $\nei{G^r}{p_{j,1}}\cap \conf{1}\ne \emptyset$, that is, $\{v\in \conf{1}\setminus \{p_{j,1}\}\mid \dist(v,p_{j,1})\le r\}\ne \emptyset$.
Here, by \Cref{prop:restrictw} and the facts that $\dist(p_{i,1}, p_{k,1})\ge r+2$ for all distinct $i,k\in [1,m]$ and that the distance from $p_{j,1}$ to every vertex contained in a variable gadget is at least $r+1$, there must be at least one vertex in the occurrence gadget corresponding to $c_j$.
Since $a_{p,j}$ is the only vertex that can be contained in $\conf{1}$ and be an $r$-neighbor of $p_{j,1}$, we must have $a_{p,j}\in \conf{1}$ for every $j\in [1,m]$.
\begin{proposition}\label{prop:restricta}
    If $a_{p,j}\in\conf{1}$, then
    $\{v^\ast_{1,j},v^\ast_{2,j},v^\ast_{3,j}\}
    \cap\conf{1}\ne\emptyset$.
\end{proposition}
\begin{proof}
    We first observe that $f_1(b_j)=b_{a,j}$; otherwise, $f_1(b_j)=b_j$ and $b_j$ is isolated in $G^r[\conf{1}]$, contradicting the $r$-connectivity of $\conf{1}$.
    Assume for a contradiction that none of the three vertices is contained in $\conf{1}$.
    We have $\nei{G^r}{p_{j,1}}\cap \conf{1}=\{a_{p,j}\}$, $\nei{G^r}{a_{p,j}}\cap \conf{1}=\{p_{j,1}, b_{a,j}\}$, and $\nei{G^r}{b_{a,j}}\cap \conf{1}=\{a_{p,j}\}$.
    Hence, $p_{j,1}$, $a_{p,j}$, and $b_{a,j}$ form a component of the graph $G^r[\conf{1}]$, contradicting the fact that $\conf{1}$ is $r$-connected.
\end{proof}
We define an assignment $\boldsymbol{x}$ by setting $x_i=\TRUE$ if $u_{i,T}\in \conf{1}$, and $x_i=\FALSE$ if $u_{i,F}\in \conf{1}$; if neither belongs to $\conf{1}$, we assign $x_i$ arbitrarily.
Note that this assignment is well-defined since there is no variable $x_i$ such that $u_{i,T}, u_{i,F}\in \conf{1}$.
Fix $j\in [1,m]$. 
By \Cref{prop:restricta} and the definition of the occurrence gadget, there is at least one variable $x_i$ that occurs in $c_j$, and the vertex in $\{u_{i,T}, u_{i,F}\}$ corresponding to the value of $x_i$ is included in $\conf{1}$.
Moreover, the corresponding literal that occurs in $c_j$ is evaluated as $\TRUE$.
Hence, every clause of $\phi$ is satisfied by $\boldsymbol{x}$; $\phi$ is satisfiable.
This concludes the proof for the only-if direction.

\subsection{Algorithmic consequences.}\label{sec:hardnessconseq}
There is a known algorithm for deciding whether a given instance admits a valid plan of length at most $1$ in polynomial time~\cite{Fekete21}.
A plan of makespan at most two is a polynomial-size certificate whose validity can be checked in polynomial time.
Thus, \Cref{thm:CUMAPFhard} establishes a sharp complexity boundary for \prbname:
the bounded-makespan variant of \prbname is polynomial-time solvable for makespan at most
one and $\NP$-complete for makespan at most two.
Equivalently, it is $\NP$-hard to compute an intermediate configuration $\mathcal Q$ such that $(S,\mathcal Q,T)$ is a valid
plan, or to correctly report that no such configuration exists.
This holds even when $\Delta(G)\leq 3$, for every integer $r\ge 2$ satisfying $r=O(|V(G)|^{1-\varepsilon})$.
The hardness result yields some consequences regarding the hardness of approximation, hardness for sublinear proximity ranges, hardness for \textsc{$r$CMAPF}, and running-time lower bounds.
\begin{corollary}
    Let $1<\rho<3/2$ and $0<\varepsilon\le 1$ be fixed constants.
    There exists a constant $c>0$ such that, unless $\P=\NP$, there is no polynomial-time $\rho$-approximation algorithm for \prbname on instances with a positive integer $r$ satisfying $1\le r=O(|V(G)|^{1-\varepsilon})$.
\end{corollary}
\begin{corollary}
    Assuming $\P\ne \NP$, there is no polynomial-time algorithm that outputs a plan for \prbname with optimal makespan, even when $r=O(\frac{D(G)}{\log D(G)})$.
\end{corollary}
\begin{proof}
    We claim that the graph $G$ constructed in our reduction has diameter $D(G)=\Omega(r\log n)$.
    Then, we can conclude that $r=O(D(G)/\log D(G))$ as follows.
    First, $D(G)\le |V|\le n^{O(1)}$; hence, $\log D(G) = O(\log n)$ holds, which implies $1/ \log n = O(1/ \log D(G))$.
    Therefore, combining this with $r=O(D(G)/ \log n)$, we obtain $r=O(D(G)/ \log D(G))$.

    We now prove the initial claim by showing that there is at least one vertex $v$ in $P_2 \coloneqq \{p_{i,2}\mid i\in [m]\}$ such that $\dist(v,p_{m,2})=\Omega(r \log n)$.
    To show this, we claim that there is a sufficiently small number of vertices in $P_2$ that have a distance at most $O(r)$ from $p_{i,2}$ for every $i\in [m]$.
    Specifically, we obtain an upper bound on the number of vertices in $P_2$ such that the distance from $u$ is at most $c r$, for every $u\in P_2$ and an arbitrary integer $c$.
    Note that each path connecting distinct $U_i$, $P_j$, the connection gadget, and the occurrence gadgets has length at least $r-1\ge r/2$, 
    and each of these gadgets is incident to a constant number of such paths (say $d$).
    Moreover, a path of length $cr$ can clearly contain at most $cr/(r/2)= 2c$ such paths.
    Hence, we can derive that the number of vertices $|\{v\in P_2\mid \dist(v,p_{m,2})\le cr\}|$ has an upper bound $\sum_{j=0}^{2c} d^j\le d^{2c+1}$.

    We now set $c=\lfloor \log_d m / 4\rfloor$.
    Then $d^{2c+1}\le d \sqrt{m}$ holds, and this can be smaller than $m$ for sufficiently large $m$.
    Hence, there is at least one vertex $v$ in $P_2$ such that $\dist(v,p_{m,2})\ge r \lceil \log_d m / 4\rceil= \Omega(r\log m)=\Omega(r\log n)$.
    This completes the proof.
\end{proof}

\begin{corollary}
    For every fixed $0<\varepsilon\le 1$, finding a plan for \textsc{$r$CMAPF} with makespan at most $2$ is \NP-hard, even when $\Delta(G)\le 3$ and $2\le r=O(|V(G)|^{1-\varepsilon})$.
\end{corollary}
\begin{proof}
    Set the target for every agent $i$ on $v\in S$ by defining the function $g_2(i)$ as $g_2(i)=f_2(v)$.
    Note that $f_2$ is a bijection; hence, this yields an instance of \textsc{$r$CMAPF}.
    The if direction of correctness clearly holds by the definition of $f_2$ and the same argument as in \Cref{sec:correctness}.
    Moreover, for the only-if direction, let $\langle \conf{0}, \conf{1}, \conf{2} \rangle$ be a valid plan for the instance, where $\conf{0}$ is the configuration corresponding to $S$, $\conf{1}$ corresponds to an intermediate (labeled) configuration, and $\conf{2}$ is a configuration such that $\conf{2}[i]=g_2(i)$.
    Here, consider the set $Q_1\coloneqq \{\conf{1}[i]\mid i\in A \}$, and the plan $\langle S, Q_1, T \rangle$. 
    This plan is a valid plan for \prbname, and hence the only-if direction of \Cref{sec:correctness} implies the only-if direction for \textsc{$r$CMAPF}.
\end{proof}

\begin{corollary}\label{cor:eth-lower-bound}
Assume that 3-\SAT on $n$ variables with $O(n)$ literals cannot be solved in $2^{o(f(n))}$ time, where $f(n)$ is non-decreasing, $f(n)=O(n)$, and $f(n)=\omega(\log n)$.
There is no $2^{o(f(c|V(G)|^\varepsilon))}$-time algorithm that computes a plan with optimal makespan, even when $\Delta(G)\leq 3$ and $r=\Theta(|V(G)|^{1-\varepsilon})$ for every fixed $0<\varepsilon\leq 1$ and some $c>0$.\footnote{A deterministic algorithm for $3$-\SAT runs in $O^\ast(1.32793^n)$ time~\cite{SAT/Liu18}, whose exponent is linear in $n$. Very recently, a preprint reported a deterministic $2^{f(n)}$-time algorithm for $3$-\SAT with $f(n)=o(n)$~\cite{SAT/OpenAI26}.}
\end{corollary}
\begin{proof}
Let $\phi$ be a Boolean formula satisfying the assumption in the corollary.
Standard transformations~\cite{SAT/Karp72,SAT/Darmann21} convert $\phi$ into an equivalent one in which
every clause contains exactly three literals and every variable occurs
exactly twice positively and exactly twice negatively.  Every formula
produced by these transformations has $O(n)$ variables and clauses.
Consequently, the latter restricted satisfiability problem on $n$
variables cannot be solved in $2^{o(f(n))}$ time either.

Apply the construction in the proof of \Cref{thm:CUMAPFhard} with
$r=\max\{2,\lceil n^{(1-\varepsilon)/\varepsilon}\rceil\}$.
The resulting graph has
$|V(G)|=\Theta(rn)=\Theta(n^{1/\varepsilon})$, and hence
$n=\Theta(|V(G)|^\varepsilon)$.
Hence, there is a constant $c>0$ such that $c|V(G)|^{\varepsilon}\le n$, where $n$ is the number of variables in the instance of 3-\SAT.
For this constant $c$, a $2^{o(f(c|V(G)|^\varepsilon))}$-time algorithm for computing an
optimal plan would decide this restricted satisfiability problem in
$2^{o(f(n))}$ time by testing whether the returned makespan is at most
$2$, contradicting the assumption.
\end{proof}

\section*{Acknowledgments}
This research was partially supported by JSPS KAKENHI Grant Number 25K21289 and JST PRESTO (JPMJPR2513).

\bibliography{ref-macro,ref}

\end{document}